\documentclass{article}

\PassOptionsToPackage{numbers}{natbib}
\usepackage[final]{neurips_2026}

\usepackage{graphicx}
\graphicspath{{figures/}}
\DeclareGraphicsExtensions{.pdf,.png,.jpg,.jpeg}
\setkeys{Gin}{draft=false}
\usepackage{wrapfig}
\usepackage{tabularx}
\usepackage[utf8]{inputenc}
\usepackage[T1]{fontenc}
\usepackage{amsmath}
\usepackage{amssymb}
\usepackage{booktabs}
\usepackage{xcolor}
\usepackage{multirow}
\usepackage{hyperref}
\usepackage{url}
\usepackage{caption}
\usepackage{subcaption}
\usepackage{array}
\usepackage{tikz}
\usepackage{float}
\usepackage{wrapfig}
\usepackage{microtype}
\usepackage{xcolor}
\usepackage{pifont}
\usepackage{natbib}
\usepackage{colortbl}

\usepackage{acro}
\acsetup{make-links=false}

\DeclareAcronym{IMU}{
  short = IMU,
  long  = Inertial Measurement Unit
}

\DeclareAcronym{PPG}{
  short = PPG,
  long  = Photoplethysmography
}

\DeclareAcronym{LoRA}{
  short = LoRA,
  long  = Low-Rank Adaptation
}

\DeclareAcronym{CNN}{
  short = CNN,
  long  = Convolutional Neural Network
}

\DeclareAcronym{MLP}{
  short = MLP,
  long  = Multi-Layer Perceptron
}

\usepackage{placeins}
\usepackage{fancyhdr}
\AtBeginDocument{\pagestyle{fancy}\fancyhf{}\fancyfoot[C]{\thepage}\fancypagestyle{plain}{\fancyhf{}\fancyfoot[C]{\thepage}}}

\newcommand{\best}[1]{\textbf{#1}}
\newcommand{\OurModel}{Mix-Token}
\newcommand{\greencheck}{\textcolor{green}{\ding{51}}}
\newcommand{\redx}{\textcolor{red}{\ding{55}}}

\title{OpenWatch: A Multimodal Benchmark for Hand Gesture Recognition on Smartwatches}

\author{
Pietro Bonazzi$^{1*}$, Youssef Ahmed$^{1*}$, Daniel Eckert$^{2}$, Andrea Ronco$^{2}$, \\ 
\textbf{Junjie Zeng$^{2}$, Dengxin Dai$^{2}$, Michele Magno$^{1}$} \\
$^{1}$ETH Z\"urich, $^{2}$Huawei Research Z\"urich \\
\small$^*$Equal contribution
}
\begin{document}
\makeatletter\renewcommand{\@noticestring}{}\makeatother
\maketitle

\begin{abstract}

Despite widespread adoption of smartwatches worldwide, open-benchmarks for wrist-based gesture recognition remain surprisingly limited. In this work, we introduce the first open-access multi-modal benchmark, \textit{OpenWatch}, for wrist-based gesture recognition using synchronized inertial and physiological sensing on a commercial smartwatch. It contains over 10 hours of \ac{IMU} and \ac{PPG} data across $50$ participants and a vocabulary of $59$ labelled gesture sequences. Furthermore, we present a subject-independent evaluation protocol including traditional and deep learning methods for time-series classification. On top of this, we develop two novel methodologies for hand-gesture recognition: (i) \textit{MixToken}, a task-specific mixture-of-experts that fuses per-channel \ac{IMU} filterbank features with cross-channel statistical tokens through learned logit mixing, and (ii) NormWear-Lora, a low-rank adaptation module for smartwatch foundation models. Our benchmarking results reveal that \ac{PPG} signals carries a substantial predictive benefit ($+12.5\%$ F1-score) for foundational smartwatch models. In addition, we show that task-specific architectures (i.e. MixToken) substantially outperforms finetuned smartwatch foundation models in terms of accuracy (F1-score=$90\%$ vs $66\%$) and memory efficiency ($223k$ vs $136M$ parameters). Finally, we also provide clear empirical guidance on the trade-offs between specialized architecture design, modality fusion, data augmentations, and foundation-model adaptation for resource-constrained wearable sensing.
\end{abstract}

% % ===============================
% 1. Introduction
% ===============================
\section{Introduction}
\begin{figure}[t]
  \centering
  \includegraphics[width=\textwidth]{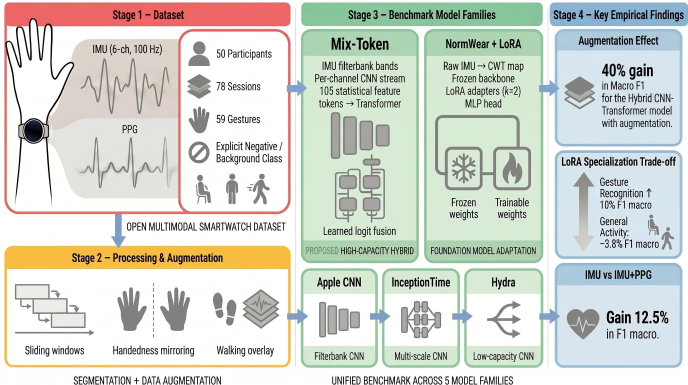}
  \caption{Overall pipeline: dataset, preprocessing and augmentation, compared model families, training protocol, and evaluation at window and clip level.}
  \label{fig:general}
\end{figure}

The rapid global adoption of smartwatches has established them as one of the most pervasive wearable sensing platforms, with more than half a billion active users worldwide in 2025~\cite{demandsage2026}. Modern devices integrate \ac{IMU} for motion tracking and \ac{PPG} sensors for cardiovascular monitoring, enabling simultaneous capture of kinematic and physiological signals from the wrist in everyday settings~\cite{castaneda2018,kim2023}. This multimodal capability has transformed smartwatches from simple fitness trackers into continuous, always-on systems supporting rich human–computer interaction, including fine-grained gesture recognition~\cite{kunwar2022, ferreira2020smartwatchReview}. Despite their ubiquity, the joint potential of these sensing modalities remains underexplored and three key gaps remain. First, public wrist-based gesture datasets are close-source, unimodal, and do not include labeled negative or background classes, limiting in-the-wild evaluation even in prior proprietary work~\cite{xuEnablingHandGesture2022b}. Second, \ac{PPG} data in commercial devices has never been adopted as informative signals linked to gesture dynamics. Third, although wearable foundation models pretrained via self-supervision~\cite{narayanswamyScalingWearableFoundation2024,pillaiPaPaGeiOpenFoundation2025,luoFoundationModelMultivariate2025} show strong transfer for coarse activities, their performance on fine-grained hand gestures remains largely unexamined.

To address these limitations, we introduce \textbf{OpenWatch}, the first open-access multimodal benchmark for smartwatch-based hand gesture recognition. OpenWatch provides synchronized six-axis \ac{IMU} and \ac{PPG} recordings from 50 participants across 78 sessions, covering 59 gesture classes with explicit negative labeling under diverse postural and activity conditions. Building on this benchmark, we perform a controlled comparison between two modeling paradigms. First, we propose \textbf{\OurModel{}}, a lightweight mixture-of-expert architecture which combines per-channel \ac{IMU} filterbank features with cross-channel statistical tokens through learned logit mixing. Second, we adapt \textbf{NormWear}~\cite{luoFoundationModelMultivariate2025}, a large wearable foundation model, using parameter-efficient \ac{LoRA}. This setup allows us to systematically study the trade-offs between task-specific architectures and pretrained foundation models for hand gesture recognition.
The main contributions of the paper are summarized as follows

\begin{itemize}
    \item \textbf{OpenWatch}: the first open smartwatch gesture dataset and benchmark with synchronized \ac{IMU} and \ac{PPG} signals, featuring the largest set of labeled positive and negative gestures.
    \item \textbf{\OurModel{}}: a lightweight mixture-of-expert architecture that integrates per-channel filterbanks with cross-channel statistical representations via learned fusion.
    \item \textbf{Empirical insights}: we show (i) that passing \ac{PPG} improves zero-shot predictions on the NormWear foundation watch model, (ii) lightweight mixture-of-expert architectures can outperform large pretrained models in both accuracy and efficiency. 
\end{itemize}

% ===============================
% 2. Related Work
% ===============================
\section{Related Work}

\subsection{Datasets for Wearable Hand Gesture Recognition}
Hand gesture recognition for wrist-worn devices has been extensively studied using a broad spectrum of sensing modalities. These include cameras~\cite{xu2015finger,yeo2019opisthenar, hanSpatioTemporalTransformerKolmogorov2025,gargConvMixFormerResourceefficientConvolution2024,althubitiDynamicGestureRecognition2024,yiEstimatingBodyHand2024}, acoustics~\cite{laput2019sensing,iravantchi2019beamband}, electromyography (EMG)~\cite{montazerinViTHGRVisionTransformerbased2022,guoSpGestureSourceFreeDomainadaptive}, radar~\cite{ahmedUWBgesturesPublicDataset2021,atzoriBuildingNinaproDatabase2012,emporioContinuousHandGesture2024}. Among them, the \ac{IMU}, has emerged as the most prevalent choice in commercial smartwatches owing to its low cost, minimal power requirements, widespread availability without line-of-sight~\cite{wen2016serendipity,kim2019imu,xu2015finger,akl2011novel, xuEnablingHandGesture2022b}.
Initial methods depended primarily on template-based techniques such as dynamic time warping (DTW)~\cite{liu2009uwave} or hidden Markov models~\cite{mckenna2004comparison}. Modern systems have shifted toward data-driven deep learning architectures, spanning convolutional networks~\cite{xuEnablingHandGesture2022b} to hybrid designs that operate on raw or filterbank-processed \ac{IMU} signals~\cite{georgi2015recognizing,iravantchi2019beamband}.
Multimodal combinations of sEMG and \ac{IMU} have demonstrated consistent gains over single-modality baselines by leveraging complementary kinematic and muscular information~\cite{shinHandGestureRecognition2024}. Wearable data glove systems with multimodal sensing (flex, force, IMU) have also been explored, achieving strong controlled-environment performance~\cite{wuDataGlovebasedGesture2023,filipowskaMachineLearningBasedGesture2024}.

Recent work has also explored \ac{PPG} signals—already embedded in commercial smartwatches—for fine-grained finger-level gesture recognition by exploiting blood-flow perturbations caused by muscle and tendon movements~\cite{zhao2018ppgFinger}. Complementary IMU-only systems on standard smartwatches have targeted thumb-to-finger interactions and broader hand gestures using patchable single six-axis sensors~\cite{anazco2021imuPatchable}. Multimodal wrist-worn approaches further combine optical/inertial sensing for touch-and-grasp detection or soft sensors for real-time hand-motion recognition, while systematic reviews highlight the growing maturity of smartwatch-based gesture methods~\cite{ferreira2020smartwatchReview,gomes2023smartwatchGestureReview}. Despite these advances, the majority existing wrist-worn datasets and models remain unimodal (or lack explicit negative/background classes), operate under controlled conditions, and have not been examined \ac{PPG} as a complementary modality for the deployment of smartwatches in-the-wild.

This paper addresses these gaps by introducing the first public benchmark that simultaneously records synchronized six-axis \ac{IMU} and \ac{PPG} data from half-hundred participants in 59 gesture classes with comprehensive negative labeling. Leveraging insights from prior IMU-focused models, we establish a benchmark comparison between a mixture-of expert lightweight architecture and large self-supervised wearable foundation models. We also provide quantitative insights into the trade-offs of modality fusion and foundation-model adaptation under a subject-independent evaluation protocol.

\subsection{Deep Learning and Wearable Foundation Models for Time-Series Gesture Recognition}

Modern approaches to wrist-based hand gesture recognition have evolved from classical template-matching and shallow statistical models toward data-driven deep learning architectures designed specifically for time-series data. Early deep learning methods relied primarily on hybrid designs that operate on raw or filterbank-processed \ac{IMU} signals~\cite{georgi2015recognizing,iravantchi2019beamband} and convolutional networks~\cite{xuEnablingHandGesture2022b}. 

More recently, self-supervised pretraining on large-scale wearable sensor time-series data has enabled general-purpose foundation models to transfer effectively to downstream tasks with minimal task-specific labels~\cite{narayanswamyScalingWearableFoundation2024,abbaspourazadLargescaleTrainingFoundation2024,pillaiPaPaGeiOpenFoundation2025,zhangSensorLMLearningLanguage2025,yuanSelfsupervisedLearningHuman2024,guTransformingLabelefficientDecoding2025,abbaspourazadWearableAccelerometerFoundation2025,erturkSensorDataFoundation2025}. Building on these advances, newer large-scale models trained on biosignals (\ac{PPG}/ECG) and customizable human activity recognition foundation models further demonstrate strong transfer potential, motivating parameter-efficient adaptation strategies (e.g., \ac{LoRA}) for fine-grained gesture tasks~\cite{qiu2025customizableHARFM}.

Among existing models, \emph{NormWear}~\cite{luoFoundationModelMultivariate2025} has emerged as a particularly strong multivariate time--frequency encoder. Pretrained on heterogeneous wearable streams (accelerometer, gyroscope, and \ac{PPG}), it is a natural candidate for smartwatch-based gesture recognition as it natively supports the sensor modalities in the OpenWatch dataset.

For a comprehensive survey of deep learning methods for time-series classification more broadly, we refer the reader to~\cite{fawazInceptionTimeFindingAlexNet2020,dempsterHYDRACompetingConvolutional2022,foumaniDeepLearningTime2023}.

% ===============================

% ===============================
\section{Methodology}

\subsection{Dataset Description}

OpenWatch\footnote{https://huggingface.co/datasets/pietrobonazzi/openwatch} includes the first multimodal dataset featuring synchronized 6-axis \ac{IMU} (100,Hz) and \ac{PPG} signals for wrist-worn gesture recognition. It was recorded from 50 participants across 78 sessions using a Huawei Smartwatch GT 4 watch with a custom data collection script connected to the activity app. A summary of the dataset characteristics is provided in Table~\ref{app:dataset_stats}.

Data was collected using a custom recording application (see Fig.~\ref{app:data_collection_interface} in Appendix) that guided participants through gesture execution with visual instructions. Each participant performed gestures under varying body postures and actions to reflect realistic usage conditions. 

Similarly to~\cite{xuEnablingHandGesture2022b}, continuous \ac{IMU} and \ac{PPG} streams were automatically segmented using a preprocessing pipeline based on bandpass filtering, motion envelope computation, smoothing, and peak detection (see preprocessing details in Appendix~\ref{app:preprocessing_pipeline}).

The final data set contains a total of 59 gesture classes, comprising both \emph{command} (positive) gestures designed to trigger device actions and a \emph{negative} (background) class representing everyday hand activities such as \emph{grabbing a cup}, \emph{typing}, \emph{waving}, and \emph{answering a phone} (full taxonomy can be found in Tables~\ref{tab:gesture_taxonomy} in Section~\ref{app:gesture_taxonomy} of the Appendix).

\begin{figure}[!htbp]
    \centering
\includegraphics[width=\columnwidth,height=0.75\textheight,keepaspectratio]{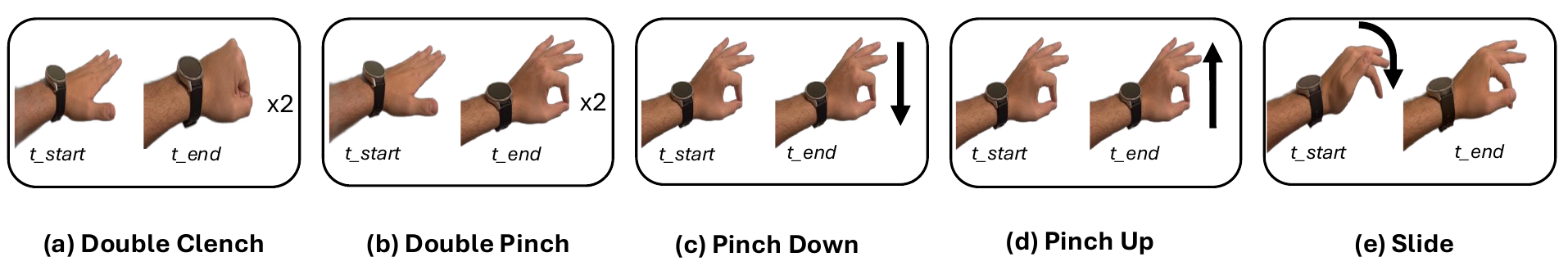}
    \caption{The five benchmark gestures: double clench, double pinch, pinch down, pinch up, and slide.}
    \label{fig:gestures_photo_examples}
\end{figure}

For these classes, we selected a subset of five positive command gestures (depicted in Fig.~\ref{fig:gestures_photo_examples}) for model evaluation and benchmark creation based on reported gesture ease, usability ratings (Appendix~\ref{app:subjective_eval}) as well as to cover different interaction primitives such as force-based actions (\emph{double\_clench}), precision finger coordination (\emph{double\_pinch}), directional variation (\emph{pinch\_up} and \emph{pinch\_down}) and continuous motion (\emph{slide}).

Table~\ref{tab:dataset_comparison} compares OpenWatch with existing smartwatch and IMU-based gesture datasets ~\cite{atzoriBuildingNinaproDatabase2012, ahmedUWBgesturesPublicDataset2021, kralikWaveGloveTransformerbasedHand2021, xuEnablingHandGesture2022b, duttaCloselyPackedStretchable2024}. 
To the best of our knowledge, OpenWatch is the first publicly available multimodal dataset to jointly capture synchronized six-axis \ac{IMU} and \ac{PPG} streams from a commercial smartwatch for gesture recognition.

\begin{table}[!htbp]
\centering
\small
\setlength{\tabcolsep}{5pt}
\begin{tabularx}{\linewidth}{@{}lXXXXXX@{}}
\toprule
\textbf{Dataset} & \textbf{Gesture Classes} & \textbf{Negative Classes} & \textbf{Number of Users} & \textbf{Sensors} & \textbf{Data Rate} & \textbf{Public} \\
\midrule
DHG-14/28~\cite{desmedtSkeletonBasedDynamicHand2016}
    & 14 & \redx & 20 & RGB & 30\,FPS & \greencheck \\
NinaPro~\cite{atzoriBuildingNinaproDatabase2012}
    & 52 & \redx & 27 & EMG & 100\,Hz & \greencheck \\
UWB-Gest.~\cite{ahmedUWBgesturesPublicDataset2021}
    & 12 & \redx & 8 & Radar & Var. & \greencheck \\
WaveGlove~\cite{kralikWaveGloveTransformerbasedHand2021}
    & 10 & \greencheck & - & \ac{IMU} & - & \greencheck \\
Apple~\cite{xuEnablingHandGesture2022b}
    & 5 & \greencheck & 512 & \ac{IMU} & 100\,Hz & \redx \\
\midrule
OpenWatch (Ours)
    & 59 & \greencheck & 50 & IMU+PPG & 100\,Hz & \greencheck \\
\bottomrule
\end{tabularx}
\caption{Comparison with representative smartwatch and IMU-based gesture datasets. OpenWatch is the first one to combine \ac{IMU} and \ac{PPG} sensors.}
\label{tab:dataset_comparison}
\end{table}

\subsection{Data Preprocessing}
Raw recordings are converted into per-gesture clips and organized by participant and condition folders. The training, validation and test splits follow a randomized selection balanced per posture and activity condition. During model training and evaluation, each clip was further processed via (i) sliding-window segmentation to increase the number of training instances and to reduce temporal sensitivity, and (ii) augmentation procedures to address watch-wrist imbalance and to introduce posture and locomotion-related variability. All time-series inputs were normalized using channel-wise z-score normalization, where mean and standard deviation were computed on the training split and then applied to validation and test data. 

\subsubsection{Window Segmentation}
Each gesture clip was segmented into multiple fixed-length windows to increase training instances and reduce temporal sensitivity. Given a clip $x \in \mathbb{R}^{T \times C}$, windows of length $W{=}100$ samples (1\,s at 100\,Hz) were extracted with stride $\Delta{=}20$ samples from a restricted sub-interval of the clip, yielding 6 overlapping windows per clip. 

\subsubsection{Data Augmentation}
Several augmentation techniques were benchmarked to improve inter-subject and in-the-wild robustness. These build upon prior work on wearable sensor data augmentation~\cite{um2017augmentation} and include three complementary procedures three complementary procedures: wrist mirroring, signal-space perturbations, and locomotion overlay.

\paragraph{Wrist mirroring.}
To compensate for the imbalance in watch-wearing wrist (37 left-wrist vs.\ 11 right-wrist), participants were duplicated with a deterministic sagittal-plane reflection applied to the inertial channels. Specifically, the signs of the X-axis accelerometer, X-axis gyroscope, and Z-axis gyroscope channels were negated to emulate execution on the opposite wrist, while all other channels remained unchanged.

\paragraph{Signal-space perturbations.}
Label-preserving perturbations were applied to windowed time series to increase tolerance to condition-dependent signal variations. These included multiplicative scaling by a random factor, monotone time-warping (including slight zooming via resampling), and additive smooth low-frequency noise. All transforms were applied before normalization.

\paragraph{Locomotion augmentation (walking overlay).}
To emulate motion artifacts encountered during natural use, we augment gesture sequences with recorded walking signals. Let $\mathbf{g}{\mathrm{acc}}(t), \mathbf{g}{\mathrm{gyro}}(t) \in \mathbb{R}^{3}$ denote the accelerometer and gyroscope channels of a gesture clip of duration $T$. We sample a walking segment $\mathbf{w}{\mathrm{acc}}(t), \mathbf{w}{\mathrm{gyro}}(t) \in \mathbb{R}^{3}$ and align it to length $T$ via cyclic tiling and cropping. The augmented signals are defined as
\begin{equation}
\label{eq:walk_aug}
\tilde{\mathbf{g}}{\mathrm{acc}}(t) = \mathbf{g}{\mathrm{acc}}(t) + \alpha,\mathbf{w}{\mathrm{acc}}(t), \quad
\tilde{\mathbf{g}}{\mathrm{gyro}}(t) = \mathbf{g}{\mathrm{gyro}}(t) + \beta,\mathbf{w}{\mathrm{gyro}}(t),
\end{equation}
where $\alpha \sim \mathcal{U}(\alpha_{\min}, \alpha_{\max})$ and $\beta \sim \mathcal{U}(\beta_{\min}, \beta_{\max})$ are independently sampled per clip. The walking signals are extracted from a continuous recording and partitioned into non-overlapping segments of length $L_w$; for $T > L_w$, we define $\mathbf{w}(t) = \mathbf{w}(t \bmod L_w)$. This augmentation injects quasi-periodic locomotion patterns and broadband perturbations that partially overlap with gesture dynamics, encouraging robustness to a dominant source of noise in unconstrained wrist-worn settings.

\subsection{Hand Gesture Recognition Benchmark}

The OpenWatch benchmark evaluates our proposed model \OurModel{} (Section~\ref{sec:mixtoken}) against three families of models: 
(1) wearable foundation models, including NormWear-Base~\cite{luoFoundationModelMultivariate2025} and its LoRA-adapted variant NormWear-LoRA (Section~\ref{sec:normwearlora}); 
(2) general time-series classification models, including InceptionTime~\cite{fawazInceptionTimeFindingAlexNet2020} and Hydra~\cite{dempsterHYDRACompetingConvolutional2022}; 
and (3) specialized hand-gesture recognition models, represented by Apple-CNN~\cite{xuEnablingHandGesture2022b}, the previous state-of-the-art on wrist-based gesture recognition.

\subsubsection{\OurModel{}}
\label{sec:mixtoken}

\OurModel{} is a lightweight and computationally efficient architecture that combines a multi-band convolutional branch to capture local temporal structure \cite{xuEnablingHandGesture2022b}, a transformer encoder \cite{vaswani2017attention} to model global statistical features and a learned fusion mechanism\cite{jacobs1991adaptive} for adaptive integration of the representation of complementary features.

\paragraph{Multi-band Convolutional Branch}

To enrich the input representation, each channel of the raw inertial signal is first decomposed using a fixed band-pass filterbank, producing multiple frequency-specific views that separate low-, mid-, and high-frequency motion dynamics. The resulting multi-channel signal is processed using a residual 1D convolutional network \cite{karim2017lstmfcnbased, xuEnablingHandGesture2022b}, which captures the local temporal structure. A global temporal pooling operation is applied at the end of the convolutional stack to obtain a fixed-dimensional embedding per channel, which is then concatenated into a unified representation and followed by a multi-layer perceptron (MLP) for the final prediction ($\hat{y}_{\mathrm{cnn}}$).

\paragraph{Statistical Transformer Branch}

In parallel, we construct a structured statistical representation derived from time-series signals, consisting of time-domain, frequency-domain, and cross-channel descriptors. These features are computed over fixed-length windows and organized into a token sequence, which is subsequently projected into an embedding space and processed by a Transformer encoder (see Appendix~\ref{app:stat_features} for a complete list). Positional encodings are added to preserve ordering over feature tokens, and a learnable query vector is used for attention-based pooling, yielding a compact representation of global statistical structure. These features are arranged as a sequence and projected into an embedding space via learned linear encoder. Positional encodings are added, and the resulting sequence is processed using a Transformer encoder~\cite{vaswani2017attention, wenTransformersTimeSeries2023}. A learnable query vector is then used for attention-based pooling and projects the feature to a compact representation of the global statistical structure. Similarly to the previous section, an MLP projects the attention vector for the final prediction ($\hat{y}_{\mathrm{atnn}}$).

\paragraph{Prediction Heads}

To combine complementary representations from convolutional, transformer-based, and joint feature spaces ($\hat{y}_{\mathrm{fused}}$), we introduce a learnable convex fusion over three prediction heads, similar to mixture-of-experts models \cite{jacobs1991adaptive}. Specifically, we compute three logits from (i) the convolutional branch, (ii) the Transformer-based statistical branch, and (iii) their concatenated representation, and combine them using normalized weights $\pi_0, \pi_1, \pi_2$.

\begin{equation}
\hat{y} = \pi_0 \hat{y}_{\mathrm{cnn}} + \pi_1 \hat{y}_{\mathrm{attn}} + \pi_2 \hat{y}_{\mathrm{fused}}, 
\quad
[\pi_0, \pi_1, \pi_2] = \mathrm{softmax}(w).
\end{equation}

In our setting, the convolution branch is known to be particularly effective for inertial hand gesture recognition \cite{xuEnablingHandGesture2022b}.
The inclusion of a separate attention-based and fusion head allows one to explicitly evaluate the standalone predictive utility of global statistical and cross-representation attention-based operators for hand gesture recognition.

\subsubsection{Foundation Model Fine-Tuning}
\label{sec:normwearlora}
We evaluate two fine-tuning strategies for the pretrained wearable foundation model NormWear~\cite{luoFoundationModelMultivariate2025}.

We first follow the protocol described in the original NormWear paper and fine-tune only a custom classification head on top of the frozen foundation model. 

In addition, we perform parameter-efficient fine-tuning using \ac{LoRA}~\cite{huLoRALowRankAdaptation2021}. The base model weights remain frozen, and only a small subset of parameters is updated: the classification head and the \ac{LoRA} modules.  This design enables efficient task adaptation while preserving pretrained representations.

The classification head is a lightweight multilayer perceptron that maps the model embedding to task logits, consisting of a LayerNorm layer followed by a linear projection to a 128-dimensional hidden space, a GELU activation, dropout regularization, and a final linear layer producing $K$ class logits.

\ac{LoRA} is applied to the query, key, value, and output projection matrices of the last two transformer blocks, as well as the [CLS]-attention fusion module. In each case, only the low-rank adaptation parameters are trained, while the original weights remain frozen.

% ===============================
% Training (shared + per-model specifics)
% ===============================
\subsubsection{Training Protocol}
\label{sec:training_optimization}

Both \OurModel{} and NormWear-\ac{LoRA} are trained to minimize a class-weighted cross-entropy loss. Inverse-frequency class weighting is applied in both cases. NormWear-LoRa additionally uses weighted sampling of minority-class windows. During training, the AdamW optimizer with decoupled weight decay is used. For \OurModel{}, a single parameter group is trained with initial learning rate $\eta_0=10^{-3}$, weight decay $\lambda=10^{-3}$, and exponential decay. For NormWear-\ac{LoRA}, the backbone is frozen and only \ac{LoRA} parameters + classifier head are optimized (separate learning rates with the same decay schedule). Both \OurModel{} and NormWear are trained with dropout regularization on the classifier head. Gradient clipping (value $1.0$ for \OurModel{}, global norm for NormWear) and mixed-precision training are employed. All models were trained for up to 100 epochs with batch size of 16 and fixed random seeds of 42. Model selection used early stopping on validation macro-F1 (patience of $5$) with checkpointing of the best epoch.

\subsubsection{Evaluation Protocol}
\label{sec:evaluation_protocol}

All models classify fixed-length windows independently in the \emph{window-level evaluation} setting. In \emph{clip-level evaluation}, a number of $k$ window predictions within each clip is aggregated for sequence-level performance. An early-stop rule assigns the clip label once any class appears in $k=3$ consecutive windows. If no class meets the threshold, majority voting is used as a fallback. We report accuracy, F1 (macro), precision and recall on all test windows.

% ===============================
% 4. Experiments and Results
% ===============================
\section{Experimentation}
\label{sec:experiments_results}

% =========================================================
\subsection{Benchmark Comparison}
\label{subsec:benchmarking}

We compare task-specific hand-gesture recognition model like Apple-CNN~\cite{xuEnablingHandGesture2022b} and \OurModel{} against wearable foundation models including NormWear-Base (frozen backbone with trained head), NormWear-LoRA (LoRA + trained head), and general time-series classification models like InceptionTime~\cite{fawazInceptionTimeFindingAlexNet2020}, Hydra~\cite{dempsterHYDRACompetingConvolutional2022}.

We report test performance at clip level (aggregation with $k=3$; Section~\ref{sec:evaluation_protocol}) using macro, micro, and weighted F1. Micro-F1 equals accuracy in this single-label multi-class setting.

\begin{table}[!htbp]
\centering
\begin{tabularx}{\textwidth}{@{}l *{7}{>{\centering\arraybackslash}X}@{}}
\toprule
\textbf{Model} &
\multicolumn{1}{c}{\textbf{Params}} &
\multicolumn{3}{c}{\textbf{Window-Level}} &
\multicolumn{3}{c}{\textbf{Clip-Level (k=3)}} \\
\cmidrule(lr){2-2}\cmidrule(lr){3-5}\cmidrule(lr){6-8}
& \textbf{Total} &
\textbf{F1} & \textbf{Acc.} & \textbf{Recall} &
\textbf{F1} & \textbf{Acc.} & \textbf{Recall} \\
\midrule
Hydra~\cite{dempsterHYDRACompetingConvolutional2022} & 74K & 0.347 & 0.584 & 0.521 & 0.391 & 0.616 & 0.529 \\
InceptionTime~\cite{fawazInceptionTimeFindingAlexNet2020} & 449K & 0.516 & 0.777 & 0.644 & 0.624 & 0.838 & 0.724 \\
Apple CNN~\cite{xuEnablingHandGesture2022b} & 51K & 0.576 & 0.774 & 0.818 & 0.756 & 0.879 & 0.949 \\ 
\rowcolor{orange!15} \textbf{\OurModel{}} & 223K & \best{0.768} & \best{0.908} & \best{0.881} & \best{0.903} & \best{0.955} & \best{0.991} \\
\midrule
NormWear-Base~\cite{luoFoundationModelMultivariate2025} & 136M & 0.430 & 0.682 & 0.617 & 0.565 & 0.788 & 0.759 \\
\rowcolor{blue!10} NormWear-LoRA & 136M & 0.560 & 0.751 & 0.806 & 0.660 & 0.803 & 0.929 \\ \midrule
NormWear-Base*~\cite{luoFoundationModelMultivariate2025} & 136M & 0.340 & 0.370 & - & 0.620 & 0.690 & - \\
\rowcolor{blue!10} NormWear-Base-PPG* & 136M & 0.370 & 0.490 & 0.415 & 0.760 & 0.840 & 0.494 \\ 
\bottomrule
\end{tabularx}
\caption{Performance comparison on window-level and clip-level aggregated predictions. * Model trained without data augmentations.} 
\label{tab:results_main}
\end{table}

\begin{figure}[!htbp]
  \centering
  \centerline{\includegraphics[width=\textwidth]{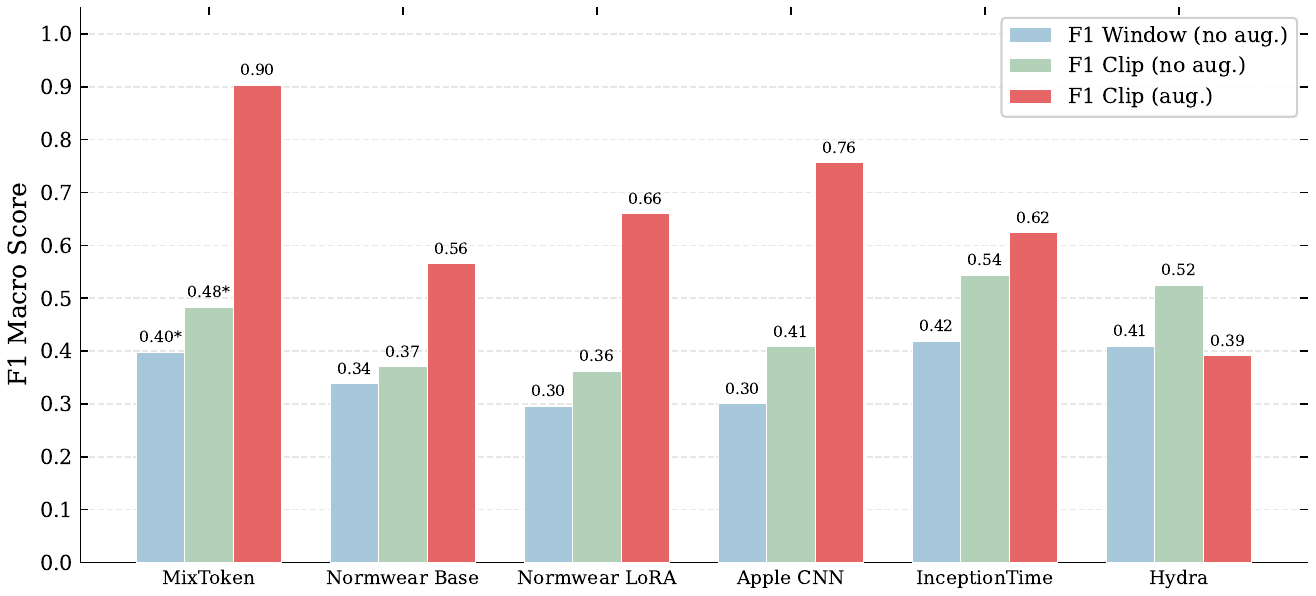}}
  \caption{Clip-level macro-F1 comparison across all models: window-level without augmentation, clip-level without augmentation, and clip-level with augmentation ($k=3$).}
  \label{fig:comprehensive_comparison}
\end{figure}

Figure~\ref{fig:comprehensive_comparison} visualizes clip-level macro-F1 across all models for window-level without augmentation, clip-level without augmentation, and clip-level with augmentation. As expected robustness increases with clip-level aggregation and data augmentation. Hydra is the one case with a negative delta which suggests that dictionary-based methods do not benefit from a larger set of synthetic data.

% =========================================================
\subsection{\OurModel{} Analysis}
\label{subsec:hybrid_analysis}

\paragraph{Fusion weight dynamics.}
\begin{wrapfigure}[16]{r}{0.38\textwidth}
  \centering
  \includegraphics[width=\linewidth]{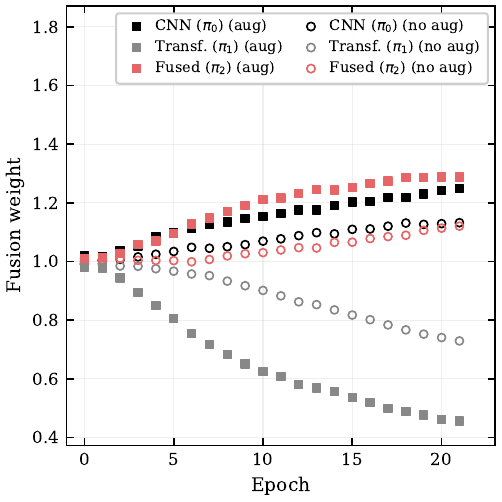}
  \caption{Fusion-weight dynamics for \OurModel{}, trained with and without data augmentation.}
  \label{fig:fusion_weights_aug_vs_noaug}
\end{wrapfigure}

As shown in Figure~\ref{fig:fusion_weights_aug_vs_noaug}, data augmentation strongly encourages synergistic use of the Transformer. This indicates that the Transformer contributes valuable complementary information, but primarily through fusion with the CNN rather than in isolation. Data augmentation creates stronger pressure for the model to learn an effective logit-mixing strategy.

Importantly, completely removing the Transformer branch causes a substantial performance drop, as shown in the feature ablation study (Figure~\ref{app:feature_ablation}), further validating that the Transformer provides meaningful complementary signal when properly integrated via the learned fusion weights.

\begin{figure}[!htbp]
  \centering
  \includegraphics[width=0.49\textwidth]{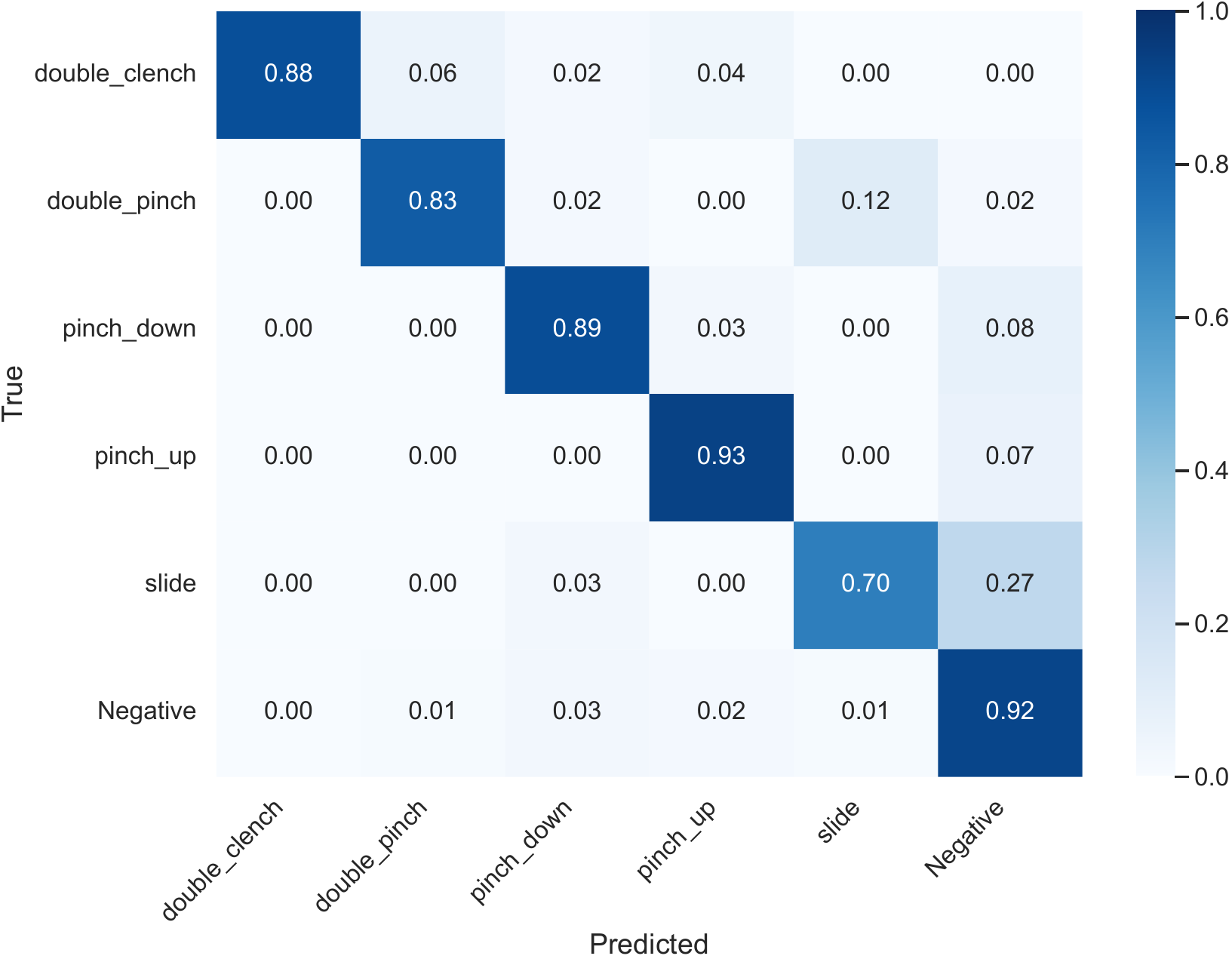}\hfill\includegraphics[width=0.49\textwidth]{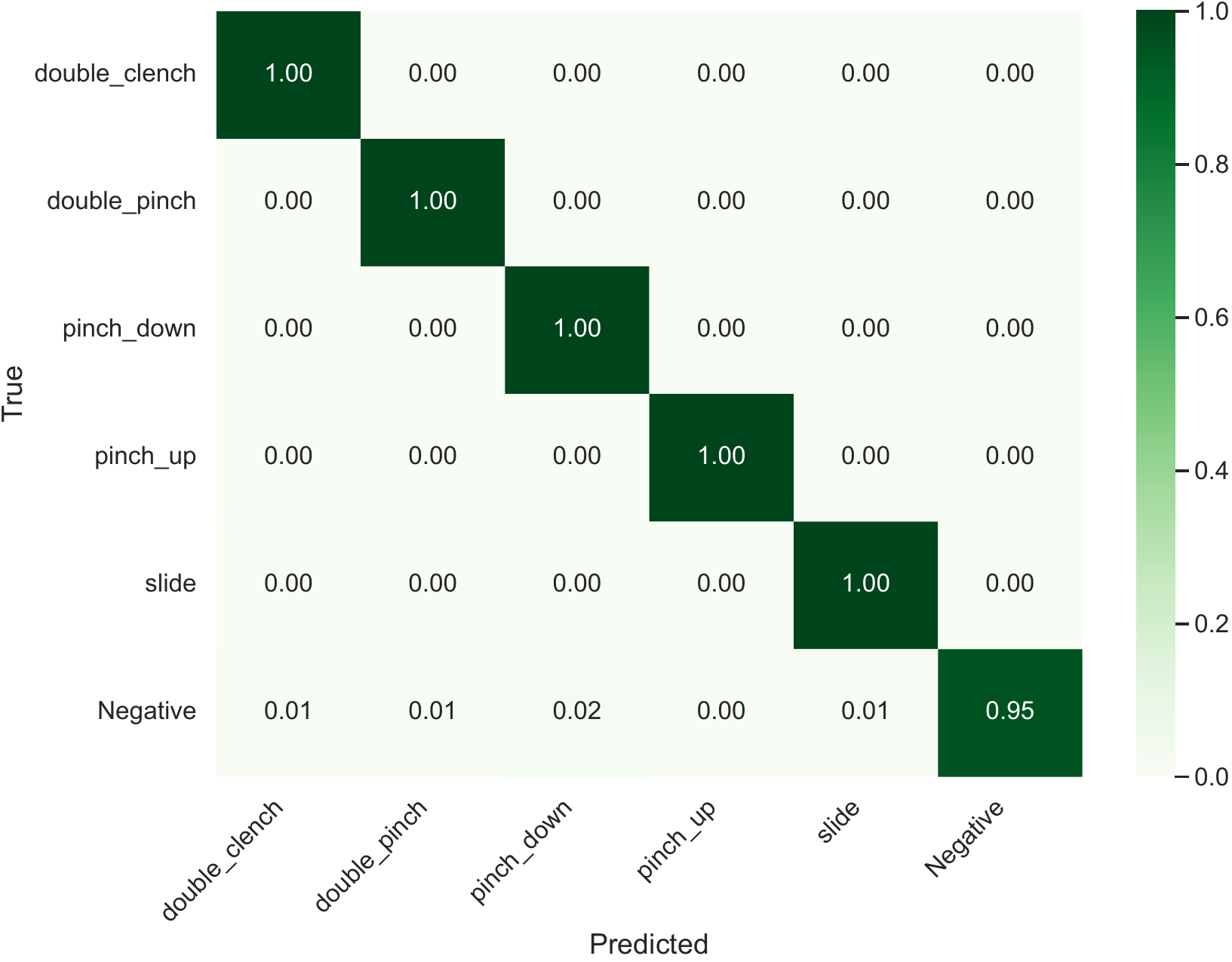}
  \caption{\OurModel{} confusion matrices (test set, with augmentation): window-level row-normalized (left) and clip-level row-normalized after aggregation with $k=3$ (right).}
  \label{fig:cm_transformer}
\end{figure}

\paragraph{Confusion matrices.}
Figure~\ref{fig:cm_transformer} presents the per-class confusion patterns for the \OurModel{}.
At window level, errors concentrate between the Negative class and \textit{pinch\_down} and \textit{slide}.
At clip level, temporal aggregation ($k=3$) largely resolves these confusions, with off-diagonal mass shifting to the Negative column, indicating that false negatives dominate over inter-gesture confusions.

\paragraph{Statistical feature ablation.}
To find the revelant set of statistical features we perform several experiments where we mask the feature and evaluate the resulting macro-F1 score. Figure~\ref{app:feature_ablation} shows the results at both window and clip level. Across the statistical components inputs, the the features of the frequency-domain are the most critical: masking them causes the largest drop ($-8.2\%$, to 0.807). Cross-channel features contribute a $-4.1\%$ drop (to 0.843). Time-domain features contribute least among the retained groups ($-1.0\%$, to 0.870). Finally, removing the entire transformer branch, yields a $-4.9\%$ drop (to 0.836), which furthers empirically supports the mixture-of-expert design of \OurModel{}. 

\subsection{NormWear Analysis}

\begin{figure*}[!htbp]
  \centering
  \includegraphics[width=\textwidth]{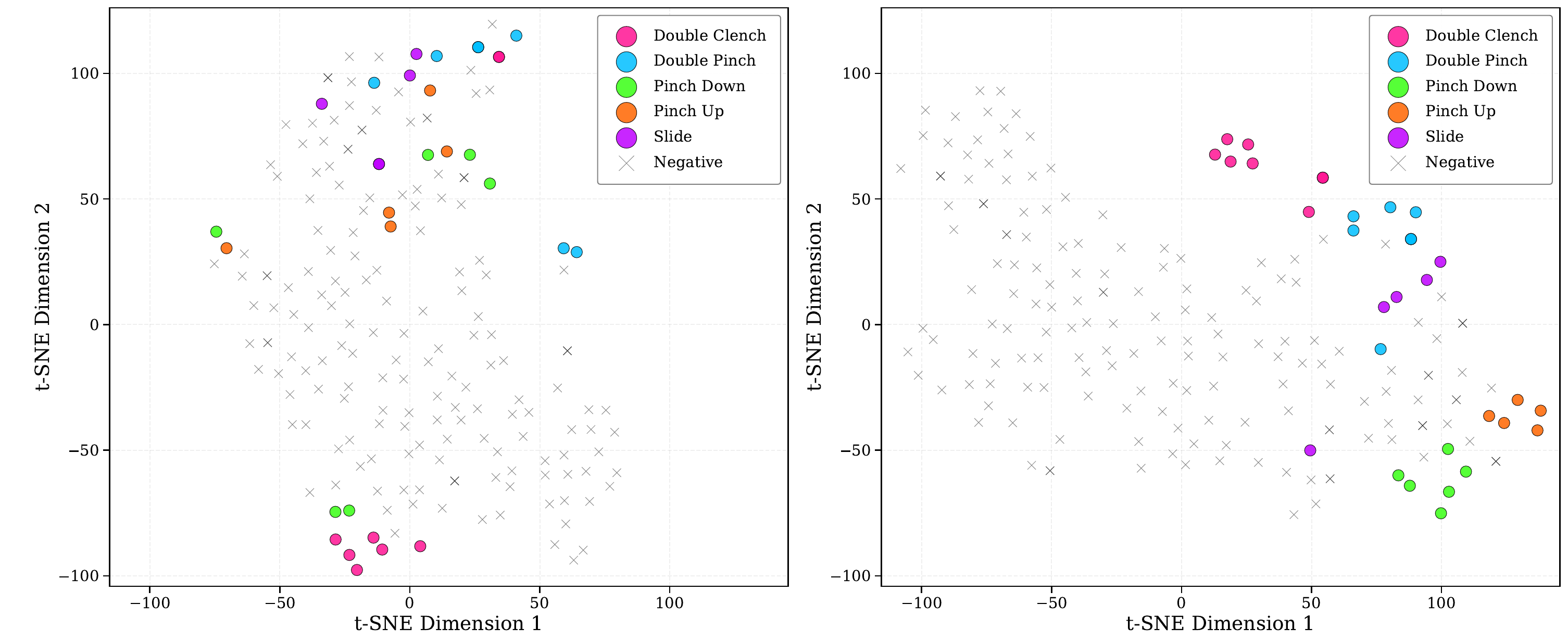}
  \caption{t-SNE visualization of clip-level embeddings (test set). Left: NormWear-Base (frozen backbone). Right: NormWear-\ac{LoRA}  (last $k=2$ encoder blocks adapted). \ac{LoRA}  fine-tuning yields more compact intra-class clusters and improved inter-class separation.}
  \label{fig:tsne_normwear}
\end{figure*}

\paragraph{NormWear-Base vs.\ -\ac{LoRA}}
NormWear-\ac{LoRA} improves over NormWear-Base by
$+0.130$ macro-F1 at window level and $+0.096$ at clip level,
confirming that adapting attention parameters in the last two
encoder blocks is sufficient to substantially improve the gesture
discrimination. The t-SNE visualization in
Figure~\ref{fig:tsne_normwear} corroborates this qualitatively:
with a frozen backbone, gesture class embeddings exhibit
substantial overlap, particularly between fine-grained pinch and
slide gestures and the Negative class, whereas after \ac{LoRA} 
adaptation embeddings show increased intra-class compactness and
clearer inter-class separation, indicating that \ac{LoRA}  reshapes the
representation space rather than merely adjusting the classifier
boundary.
\vspace{-8pt}

Figure~\ref{app:cm_normwear_lora} in the Appendix shows the confusion metrics obtained. Similar to results for other models, clip-level aggregation improves accuracy.

\paragraph{Impact of \ac{LoRA} on activity recognition.}

Gesture-specific fine-tuning of NormWear with \ac{LoRA} leads to a modest degradation in general activity recognition performance on the UCI-HAR dataset ($-3.8$ percentage points in macro-F1). As shown in Figure~\ref{app:ucihar_linear_probe}, this drop is concentrated among the static posture classes (sitting, standing, and laying), where the model’s embeddings become slightly less separable after adaptation. In contrast, performance on dynamic locomotion activities (walking, walking upstairs, and walking downstairs) remains largely unaffected or even slightly improved. 

% =========================================================
\paragraph{NormWear-Base Finetuning with PPG}
\label{subsec:ppg}

We compare NormWear-Base in IMU-only and IMU+PPG configurations by finetuning the model on the non-augmented dataset. 

\begin{wrapfigure}[15]{r}{0.5\textwidth}
  \centering
  \includegraphics[width=\linewidth]{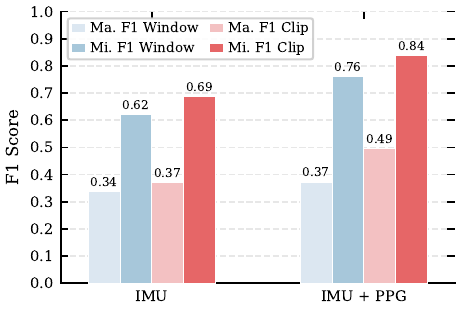}
  \caption{Effect of \ac{PPG} on NormWear-Base (no augmentation). Adding \ac{PPG} yields a modest window-level gain but a $3.6\times$ larger gain at clip level, consistent with a temporally stabilizing role..}
  \label{fig:imu_ppg_comparison}
\end{wrapfigure}

Adding \ac{PPG} improves window-level macro-F1 by $+0.035$ and clip-level by $+0.125$ ($3.6\times$ ratio), suggesting that \ac{PPG} provides a useful signal to predict gestures in wearable foundation models. This is consistent with prior findings on the value of multi-wavelength PPG for mitigating motion artifacts in wearable settings~\cite{lee2020ppg}.

% ===============================
% 6. Limitations
% ===============================
\section{Limitations}

While OpenWatch and Mix-Token provide the first open multimodal benchmark for commercial smartwatch-based gesture recognition and state-of-the-art results, several limitations should be acknowledged.

First, although the dataset defines a rich taxonomy of 59 gesture classes (including both positive and negative/background activities), the core benchmark focuses on a small subset of five representative command gestures. This design choice is intentional: rather than attempting to classify an exhaustive and fine-grained gesture vocabulary—which is likely unrealistic given the sensing limitations of IMU and PPG—we prioritize a set of gestures that are practical, distinguishable, and suitable for real-world interaction. Second, the dataset comprises recordings from 50 participants under semi-structured conditions. Despite efforts to introduce variability through different postures and augmentation strategies, the data may not fully capture the diversity of real-world usage, including long-term deployment effects, sensor drift, or unseen motion patterns. Third, although we adopt a subject-independent evaluation protocol, we do not explore personalization or user adaptation, which are important for practical deployment in wearable systems. Addressing these limitations represents an important direction for future work.

% ===============================
% 7. Conclusion
% ===============================
\section{Conclusion}

In this work, we introduced \textbf{OpenWatch}, the first open-access multimodal benchmark for smartwatch-based hand gesture recognition. By releasing over 10 hours of synchronized IMU and PPG recordings from 50 participants across 59 gesture classes—including explicit negative/background activities—we provide the community with a realistic, subject-independent evaluation framework that addresses critical gaps in existing datasets. Our comprehensive benchmarking reveals several actionable insights for resource-constrained wearable sensing. First, PPG signals, already present in commercial smartwatches, deliver substantial predictive value for foundation models (+12.5\% clip-level macro F1), acting as a temporally stabilizing modality that complements IMU motion cues. Second, we designed lightweight task-specific architecture, such as \textbf{MixToken} (223K parameters), which substantially outperforms both large pretrained foundation models (136M parameters) and prior specialized CNNs. \textbf{MixToken} achieves 90.3\% clip-level macro F1 while maintaining orders-of-magnitude better memory and compute efficiency. This demonstrates that domain- and task-specific inductive biases remain highly valuable even in the era of foundation models. Finally, our controlled experiments clarify the practical trade-offs between modality fusion, data augmentation strategies, statistical feature design, and parameter-efficient adaptation. OpenWatch and MixToken together establish a strong new baseline for on-device hand gesture recognition and open promising avenues for future research. Immediate opportunities include expanding the gesture vocabulary and participant diversity, exploring more advanced multimodal fusion techniques, investigating personalization under strict privacy constraints, and integrating these capabilities into end-to-end smartwatch interaction systems. This benchmark and the accompanying empirical guidance will accelerate progress toward intuitive, always-available gesture interfaces on everyday wearable devices, ultimately making human–computer interaction more natural and accessible in real-world settings.

\section*{Ethics Statement}
 Data collection was conducted under informed consent from all participants. Participants were briefed on the study objectives, the types of sensor data recorded, and their right to withdraw at any time without consequence. Collected sensor data were pseudonymized prior to analysis and public release; no personally identifiable information is included in the dataset. Participants were not financially compensated. 

\section*{NeurIPS Paper Checklist}

\begin{enumerate}

\item {\bf Claims}
    \item[] Question: Do the main claims made in the abstract and introduction accurately reflect the paper's contributions and scope?
    \item[] Answer: \answerYes{} % Replace by \answerYes{}, \answerNo{}, or \answerNA{}.
    \item[] Justification: The abstract and introduction clearly state the key contributions—namely the OpenWatch dataset and benchmark, the Mix-Token architecture, and the empirical findings—and these are directly supported by the experimental results and analyses presented in the paper. The claims are specific, aligned with the reported benchmarks, and do not overstate generalization beyond the evaluated setting.

\item {\bf Limitations}
    \item[] Question: Does the paper discuss the limitations of the work performed by the authors?
    \item[] Answer: \answerYes{} % Replace by \answerYes{}, \answerNo{}, or \answerNA{}.
    \item[] Justification: The paper includes a dedicated limitations section discussing the gesture subset, semi-structured data collection, and lack of personalization, clearly outlining future directions for real-world deployment.

\item {\bf Theory assumptions and proofs}
    \item[] Question: For each theoretical result, does the paper provide the full set of assumptions and a complete (and correct) proof?
    \item[] Answer: \answerNA{} % Replace by \answerYes{}, \answerNo{}, or \answerNA{}.
    \item[] Justification: The paper does not present formal theoretical results requiring assumptions or proofs, focusing instead on empirical benchmarking and model design.

    \item {\bf Experimental result reproducibility}
    \item[] Question: Does the paper fully disclose all the information needed to reproduce the main experimental results of the paper to the extent that it affects the main claims and/or conclusions of the paper (regardless of whether the code and data are provided or not)?
    \item[] Answer: \answerYes{} % Replace by \answerYes{}, \answerNo{}, or \answerNA{}.
    \item[] Justification: The paper provides detailed descriptions of dataset construction, preprocessing, model architectures, and training procedures sufficient to reproduce the main experimental results.

\item {\bf Open access to data and code}
    \item[] Question: Does the paper provide open access to the data and code, with sufficient instructions to faithfully reproduce the main experimental results, as described in supplemental material?
    \item[] Answer: \answerYes{} % Replace by \answerYes{}, \answerNo{}, or \answerNA{}.
    \item[] Justification: The dataset and benchmark are introduced as open-access resources, and the paper describes their structure and intended public release for reproducibility.

\item {\bf Experimental setting/details}
    \item[] Question: Does the paper specify all the training and test details (e.g., data splits, hyperparameters, how they were chosen, type of optimizer) necessary to understand the results?
    \item[] Answer: \answerYes{} % Replace by \answerYes{}, \answerNo{}, or \answerNA{}.
    \item[] Justification: The paper specifies training protocols, hyperparameters, architectures, preprocessing steps, and evaluation procedures. In addition, the entire benchmark is open-access thus enabling clear understanding of architectures and preprocessing steps.

\item {\bf Experiment statistical significance}
    \item[] Question: Does the paper report error bars suitably and correctly defined or other appropriate information about the statistical significance of the experiments?
    \item[] Answer: \answerYes{} % Replace by \answerYes{}, \answerNo{}, or \answerNA{}.
    \item[] Justification: The paper reports point estimates of performance metrics and includes error bars and confusion met and several ablation studies.

\item {\bf Experiments compute resources}
    \item[] Question: For each experiment, does the paper provide sufficient information on the computer resources (type of compute workers, memory, time of execution) needed to reproduce the experiments?
    \item[] Answer: \answerNo{} % Replace by \answerYes{}, \answerNo{}, or \answerNA{}.
    \item[] Justification: The paper reports training hyperparameters (epochs, batch size, optimizer settings) and total training duration, which allow estimation of compute requirements on standard hardware, but does not specify exact compute worker type or memory usage.
    
\item {\bf Code of ethics}
    \item[] Question: Does the research conducted in the paper conform, in every respect, with the NeurIPS Code of Ethics \url{https://neurips.cc/public/EthicsGuidelines}?
    \item[] Answer: \answerYes{} % Replace by \answerYes{}, \answerNo{}, or \answerNA{}.
    \item[] Justification: The study follows ethical guidelines with informed consent, pseudonymization of data, and transparency about data collection procedures.

\item {\bf Broader impacts}
    \item[] Question: Does the paper discuss both potential positive societal impacts and negative societal impacts of the work performed?
    \item[] Answer: \answerNo{} % Replace by \answerYes{}, \answerNo{}, or \answerNA{}.
    \item[] Justification: The paper does not explicitly discuss broader societal impacts, although potential implications (e.g., wearable interaction systems) are implicit in the application domain.
    
\item {\bf Safeguards}
    \item[] Question: Does the paper describe safeguards that have been put in place for responsible release of data or models that have a high risk for misuse (e.g., pre-trained language models, image generators, or scraped datasets)?
    \item[] Answer: \answerNo{} % Replace by \answerYes{}, \answerNo{}, or \answerNA{}.
    \item[] Justification: The dataset does not pose high misuse risks (e.g., no sensitive personal data or generative models), so additional safeguards are not required.

\item {\bf Licenses for existing assets}
    \item[] Question: Are the creators or original owners of assets (e.g., code, data, models), used in the paper, properly credited and are the license and terms of use explicitly mentioned and properly respected?
    \item[] Answer: \answerNo{} % Replace by \answerYes{}, \answerNo{}, or \answerNA{}.
    \item[] Justification: While prior datasets and models are cited, the paper does not explicitly specify licenses or terms of use for all referenced assets.

\item {\bf New assets}
    \item[] Question: Are new assets introduced in the paper well documented and is the documentation provided alongside the assets?
    \item[] Answer: \answerYes{} % Replace by \answerYes{}, \answerNo{}, or \answerNA{}.
    \item[] Justification: The paper introduces the OpenWatch dataset and provides detailed documentation of data collection, preprocessing, and structure.

\item {\bf Crowdsourcing and research with human subjects}
    \item[] Question: For crowdsourcing experiments and research with human subjects, does the paper include the full text of instructions given to participants and screenshots, if applicable, as well as details about compensation (if any)? 
    \item[] Answer: \answerNo{} % Replace by \answerYes{}, \answerNo{}, or \answerNA{}.
    \item[] Justification: The paper describes the human-subject data collection protocol and includes the data acquisition interface and instructions (Figure\ref{app:data_collection_interface}).

\item {\bf Institutional review board (IRB) approvals or equivalent for research with human subjects}
    \item[] Question: Does the paper describe potential risks incurred by study participants, whether such risks were disclosed to the subjects, and whether Institutional Review Board (IRB) approvals (or an equivalent approval/review based on the requirements of your country or institution) were obtained?
    \item[] Answer: \answerNo{} % Replace by \answerYes{}, \answerNo{}, or \answerNA{}.
    \item[] Justification: : The paper mentions informed consent and ethical procedures but does not explicitly state IRB or equivalent approval.

\item {\bf Declaration of LLM usage}
    \item[] Question: Does the paper describe the usage of LLMs if it is an important, original, or non-standard component of the core methods in this research? Note that if the LLM is used only for writing, editing, or formatting purposes and does \emph{not} impact the core methodology, scientific rigor, or originality of the research, declaration is not required.
    %this research? 
    \item[] Answer: \answerNo{} % Replace by \answerYes{}, \answerNo{}, or \answerNA{}.
    \item[] Justification: Large language models are not part of the methodology or experimental pipeline.

\end{enumerate}

% ===============================
% Bibliography
% ===============================
\bibliographystyle{plainnat}
\bibliography{bib_list}

% ===============================
% Appendix
% ===============================
\appendix

\section{Dataset}

\subsection{Dataset Characteristics}

The Table~\ref{app:dataset_stats} summarizes the key characteristics of the proposed smartwatch gesture dataset, including participant demographics, recording conditions, and sensor modalities. The dataset comprises 50 participants with a diverse age distribution and varying recording setups (e.g., posture and wrist placement). In total, 78 sessions were collected, covering 59 gesture classes (including a negative/background class). Sensor data consist of high-frequency IMU signals (100 Hz) and photoplethysmography (PPG), enabling both motion and physiological analysis.

\begin{table}[!htbp]
\centering
\small
\setlength{\tabcolsep}{4pt}
\renewcommand{\arraystretch}{1.08}
\begin{tabular}{@{}p{0.36\textwidth} p{0.56\textwidth}@{}}
\toprule
\textbf{Factor} & \textbf{Information} \\
\midrule
\multicolumn{2}{@{}l}{\textit{Demographics}} \\
Total participants & 50 \\
Self-identified gender & Male: 41; Female: 5; Not specified: 4 \\
Age (years) & Min: 21; Max: 51; Mean: 27.2 $\pm$ 6.0 \\
Watch wrist & Left: 37; Right: 11; Not specified: 2 \\
\midrule
\multicolumn{2}{@{}l}{\textit{Gesture Data}} \\
Body posture & Sitting (39); Standing (25); Arm down (10); Walking (4) \\
Total sessions & 78 sessions across 50 participants \\
Gesture classes & 59 (including labeled negative class) \\
\midrule
\multicolumn{2}{@{}l}{\textit{Sensors}} \\
IMU sampling rate & 100 Hz \\
Physiological signal & Photoplethysmography (PPG) \\
\bottomrule
\end{tabular}
\caption{Summary of the Proposed Smartwatch Gesture Dataset}
\label{app:dataset_stats}
\end{table}

\subsection{Gesture Taxonomy}
\label{app:gesture_taxonomy}

Table~\ref{tab:gesture_taxonomy} lists the 27 positive gesture classes (IDs 0--26) and negative classes used in the dataset. For benchmarking, the five evaluated gestures are: \textit{double\_clench} (ID~3), \textit{double\_pinch} (ID~4), \textit{pinch\_up} (ID~16), \textit{pinch\_down} (ID~13), and \textit{slide} (ID~20). Video demonstrations of all 59 gesture classes are available at \url{https://www.youtube.com/playlist?list=PLSQoDCloQ92WnNrkKAOZqsaCWkw3cmfLp}.

\begin{table*}[t]
\centering
\footnotesize
\setlength{\tabcolsep}{4pt}
\begin{tabular*}{\textwidth}{@{\extracolsep{\fill}} r l r l r l r l}
\toprule
\textbf{ID} & \textbf{Name} & \textbf{ID} & \textbf{Name} & \textbf{ID} & \textbf{Name} & \textbf{ID} & \textbf{Name} \\
\midrule
\multicolumn{8}{l}{\textit{Positive gestures}} \\
\midrule
0  & clench            & 7  & double spread     & 14 & pinch left        & 21 & snap \\
1  & deviate in        & 8  & extend            & 15 & pinch right       & 22 & spiderman \\
2  & deviate out       & 9  & flex              & 16 & \textbf{pinch up}   & 23 & spread \\
3  & \textbf{double clench} & 10 & index pointing & 17 & pinky pinch      & 24 & thumbs down \\
4  & \textbf{double pinch}  & 11 & peace          & 18 & rotate in        & 25 & thumbs up \\
5  & double pinky pinch & 12 & pinch           & 19 & rotate out       & 26 & vulcan salute \\
6  & double snap       & 13 & \textbf{pinch down} & 20 & \textbf{slide}  &    & \\
\midrule
\multicolumn{8}{l}{\textit{Negative gestures (background)}} \\
\midrule
27 & grab cup        & 35 & no waving       & 43 & handshake        & 51 & finger up \\
28 & type phone      & 36 & money sign      & 44 & open palm        & 52 & peace up \\
29 & type computer   & 37 & love fingers    & 45 & answer phone     & 53 & pistol gun \\
30 & wash hands      & 38 & waving hello    & 46 & ok sign          & 54 & cat grab \\
31 & waiting bored   & 39 & knock on wood   & 47 & power sign       & 55 & love hands \\
32 & write pen       & 40 & italian pinch   & 48 & small amount     & 56 & good luck \\
33 & cheering fist   & 41 & clapping        & 49 & steepling        & 57 & air \\
34 & stop sign       & 42 & horns           & 50 & bill please      & 58 & balling \\
\bottomrule
\end{tabular*}
\caption{Gesture taxonomy. Positive classes (IDs 0--26) include evaluated gestures in \textbf{bold}. Negative classes (IDs 27--58) are collapsed into the background class.}
\label{tab:gesture_taxonomy}
\end{table*}

\subsection{Dataset Collection}

\begin{figure*}[!htbp]
    \centering
    \includegraphics[width=\columnwidth]{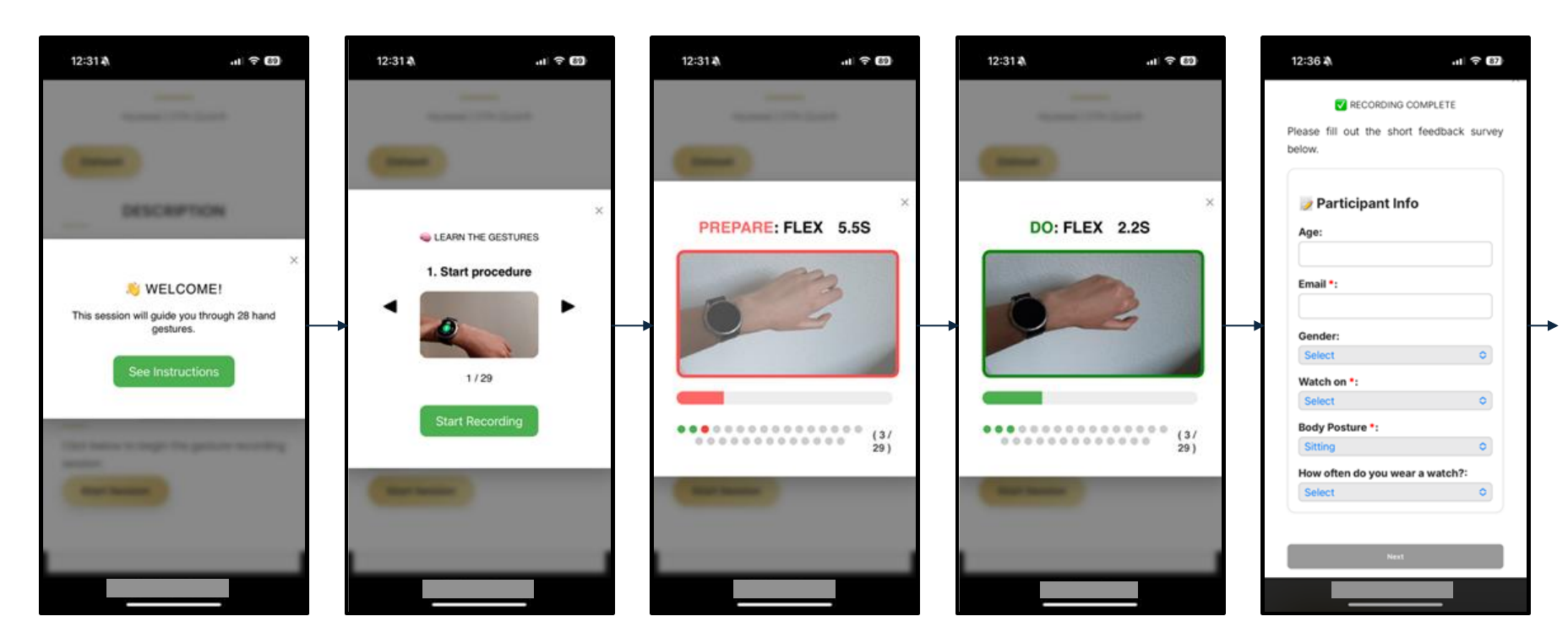}\\[1pt]
    \includegraphics[width=\columnwidth]{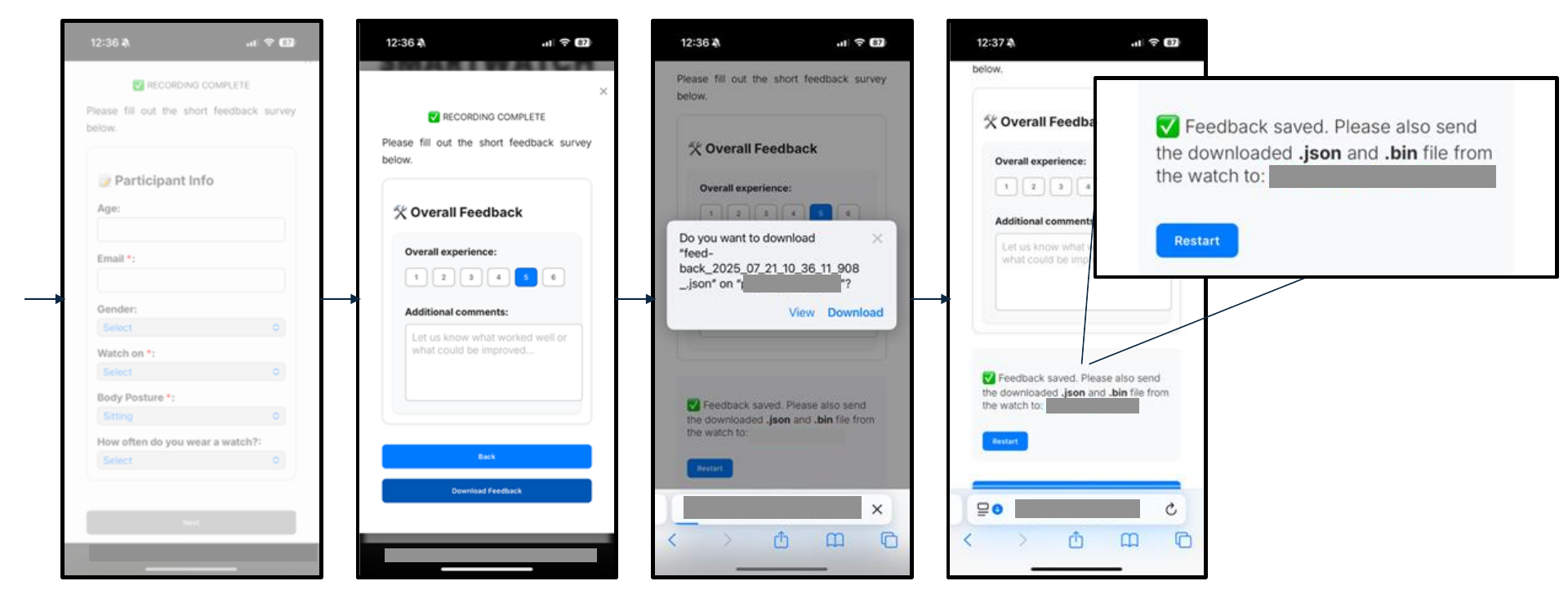}
    \caption{Data collection interface. Top: gesture selection, instruction display with countdown timer (prepare/do phases), and progress tracking. Bottom: post-session feedback survey collecting participant demographics, subjective ratings, and data export.}
    \label{app:data_collection_interface}
\end{figure*}

  The Figure~\ref{app:data_collection_interface} illustrates the custom data acquisition interface used during recording, which is publicly available at                                       
  \url{https://pietrobonazzi.com/projects/huawei}. Participants viewed the interface on their smartphones, which provided gesture prompts and countdown timers while leaving them free to move 
  naturally during motion-based body postures (e.g., walking). The top panel shows the real-time interaction flow, including gesture prompts, countdown timers for the preparation and
  execution phases, and session progress tracking. The bottom panel presents the post-session survey used to collect demographic information, subjective feedback, and facilitate data export. 
  The interface supported multiple parallel recording sessions: some participants shared the same on-screen instructions while up to four smartwatches recorded simultaneously, enabling efficient
  multi-device data collection in a single pass.

\subsection{Data Preprocessing}

Continuous sensor streams were segmented into gesture instances using an
automatic segmentation pipeline adapted from
Xu~et~al.~\cite{xuEnablingHandGesture2022b}. The pipeline, 
%illustrated in Fig.~\ref{app:preprocessing_pipeline},
proceeds in four stages:
(a)~a bandpass filter is applied to the raw 6-axis \ac{IMU} signals to remove DC offset and high-frequency noise;
(b)~the L2 magnitude of the filtered accelerometer and gyroscope channels is computed and summed into a single total-motion envelope; 
(c)~a moving-average kernel smooths the envelope to suppress transient fluctuations; and
(d)~peak detection on the smoothed signal identifies candidate gesture centres, from which fixed-length windows are extracted to define gesture boundaries.
This procedure enables consistent segmentation across sessions while minimising manual intervention.

Additional processing steps include time-drift correction and alignment between sensor streams, as well as consistency checks to verify temporal coherence across modalities. Dataset statistics, label mappings, and quality-control summaries were generated programmatically as part of the preprocessing pipeline.  Multi-channel \ac{PPG} values were averaged before pre-processing.

% \begin{figure}[htbp]
%     \centering
%     \includegraphics[width=\columnwidth]{preprocessing1.pdf}\\[2pt]
%     \includegraphics[width=\columnwidth]{preprocessing2.pdf}\\[2pt]
%     \includegraphics[width=\columnwidth]{preprocessing3.pdf}
%     \caption{Automatic segmentation pipeline: (a) bandpass filtering of raw \ac{IMU} signals, (b) accelerometer and gyroscope magnitude computation, (c) moving-average smoothing of the total motion signal, and (d) peak detection on the smoothed envelope.}
%     \label{app:preprocessing_pipeline}
% \end{figure}

\subsection{Subjective Gesture Evaluation}

\begin{figure}[!htbp]
  \centering
  \includegraphics[width=\textwidth]{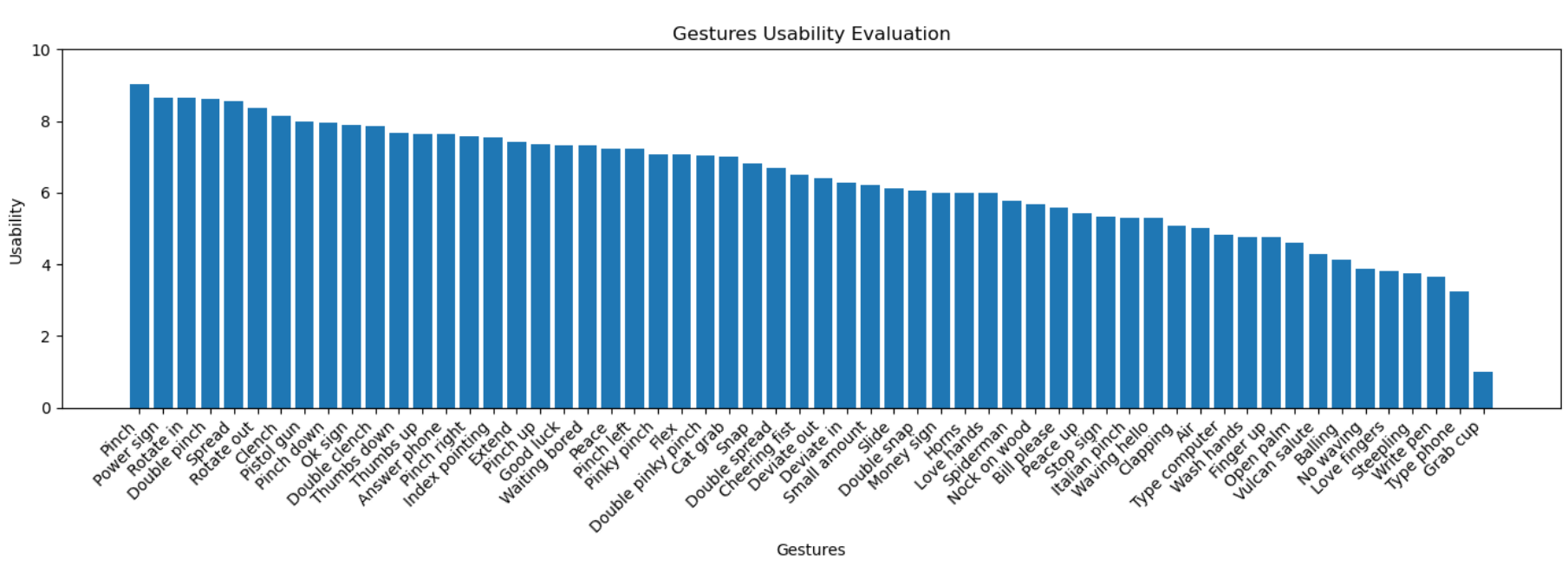}
  \vspace{2pt}
  \includegraphics[width=\textwidth]{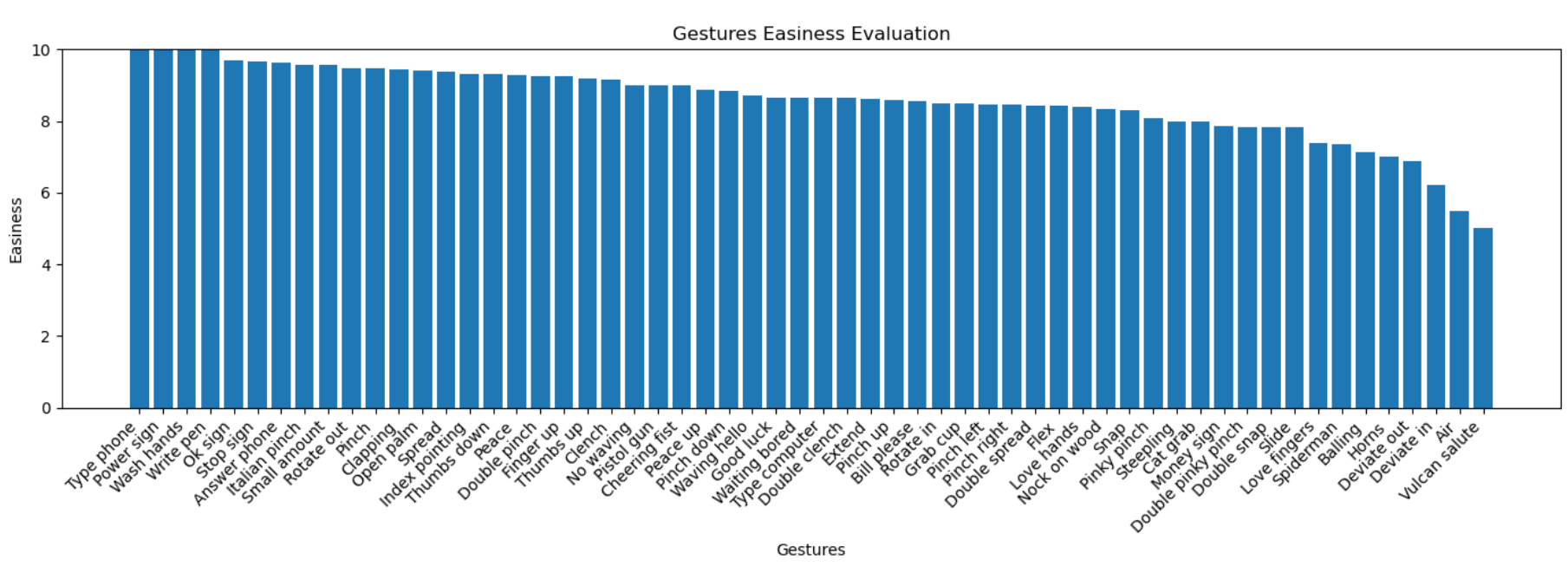}
  \caption{Subjective evaluation of gesture quality. Top: mean usability scores per gesture. Bottom: mean easiness scores per gesture. Higher scores indicate greater user comfort and suitability for smartwatch interaction.}
  \label{app:subjective_eval}
\end{figure}

Figure~\ref{app:subjective_eval} reports the mean usability and easiness scores collected from participants during data acquisition. Gestures selected for benchmarking (\textit{double\_clench}, \textit{double\_pinch}, \textit{pinch\_up}, \textit{pinch\_down}, \textit{slide}) consistently ranked among the highest in both dimensions.

\section{Methodology}

\subsection{Statistical Encoder Features}

The Table~\ref{app:stat_features} details the statistical features used in the Transformer branch, grouped into time-domain, frequency-domain, and cross-channel descriptors. These features capture complementary aspects of the sensor signals and are aggregated into token sequences for downstream modeling.

\begin{table}[!htbp]
\centering
\small
\begin{tabular}{p{0.22\textwidth} p{0.72\textwidth}}
\hline
\textbf{Feature Group} & \textbf{Features} \\
\hline

Time-domain &
Mean, variance, standard deviation, min, max, range, median, 25\% and 75\% quantiles, interquartile range, RMS, median absolute deviation, energy, zero-crossing rate, autocorrelation (lag 1, 2), skewness, kurtosis, Hjorth activity, mobility, complexity \\

Frequency-domain &
Spectral centroid, spectral entropy, spectral flux, peak frequency, low-band energy, mid-band energy, high-band energy, spectral roll-off (85\%), spectral flatness, top-$k$ dominant frequencies \\

Cross-channel &
Pearson correlation, Spearman correlation, cosine similarity, mutual information, PCA explained variance ratios \\
\hline
\end{tabular}
\caption{Statistical feature groups used in the Transformer branch. Features are computed per channel or across channels depending on the descriptor type and aggregated into a token sequence for Transformer encoding.}
\label{app:stat_features}
\end{table}

\begin{figure}[!htbp]                                                                                                                                                       
\centering                                                                                                                                                                                 
\includegraphics[width=0.6\textwidth]{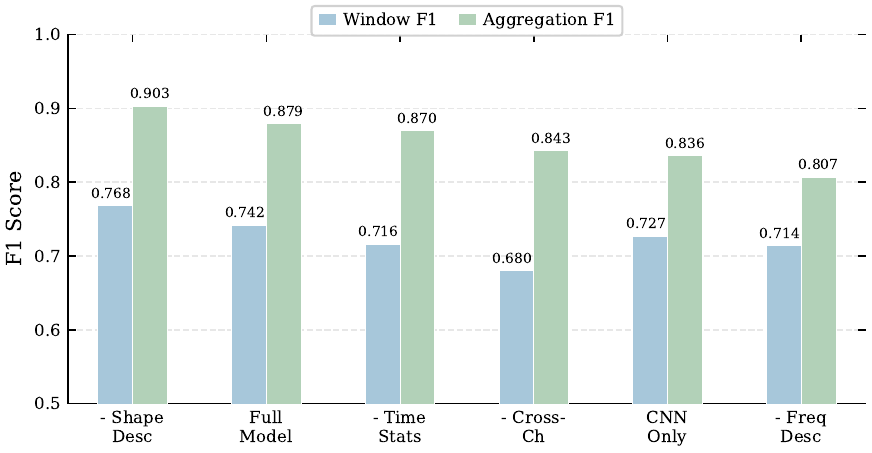}                                                                                                              
\caption{Feature group masking ablation for the \OurModel{} (with augmentation). Each group of bars shows window-level and clip-level macro-F1 when one component is masked from the full  
model (clip-level baseline: 0.879). Masking shape descriptors \emph{improves} performance to 0.903; all other components contribute positively, with frequency-domain features showing the   
largest impact.}                                                                                                                                                                             
\label{app:feature_ablation}                                                                                                                                                               
\end{figure}

Figure~\ref{app:feature_ablation} evaluates the contribution of different feature groups by masking them individually. Results show that frequency-domain features provide the largest performance gain, while removing certain shape descriptors can slightly improve performance, suggesting potential redundancy.

% \subsection{NormWear-LoRA architecture}

% This Figure~\ref{app:arch_normwear_lora} presents the adaptation pipeline for the pretrained NormWear model. IMU signals are first transformed into time–frequency representations, then processed by a frozen backbone. \ac{LoRA} is applied to the final layers, while a lightweight classification head is trained for gesture recognition.

% \begin{figure}[!htbp]
%   \centering
%   \includegraphics[width=\textwidth]{figures/normwear_lora_arch_new.pdf}
%   \caption{NormWear adaptation pipeline. Raw \ac{IMU} windows are converted to a time--frequency representation and encoded by a frozen NormWear backbone; LoRA updates are applied to the last $k$ blocks and a lightweight ResidualMLP head is trained for gesture classification.}
%   \label{app:arch_normwear_lora}
% \end{figure}

\subsection{NormWear-LoRA architecture}

\begin{figure}[!htbp]
  \centering
  \includegraphics[width=0.49\linewidth]{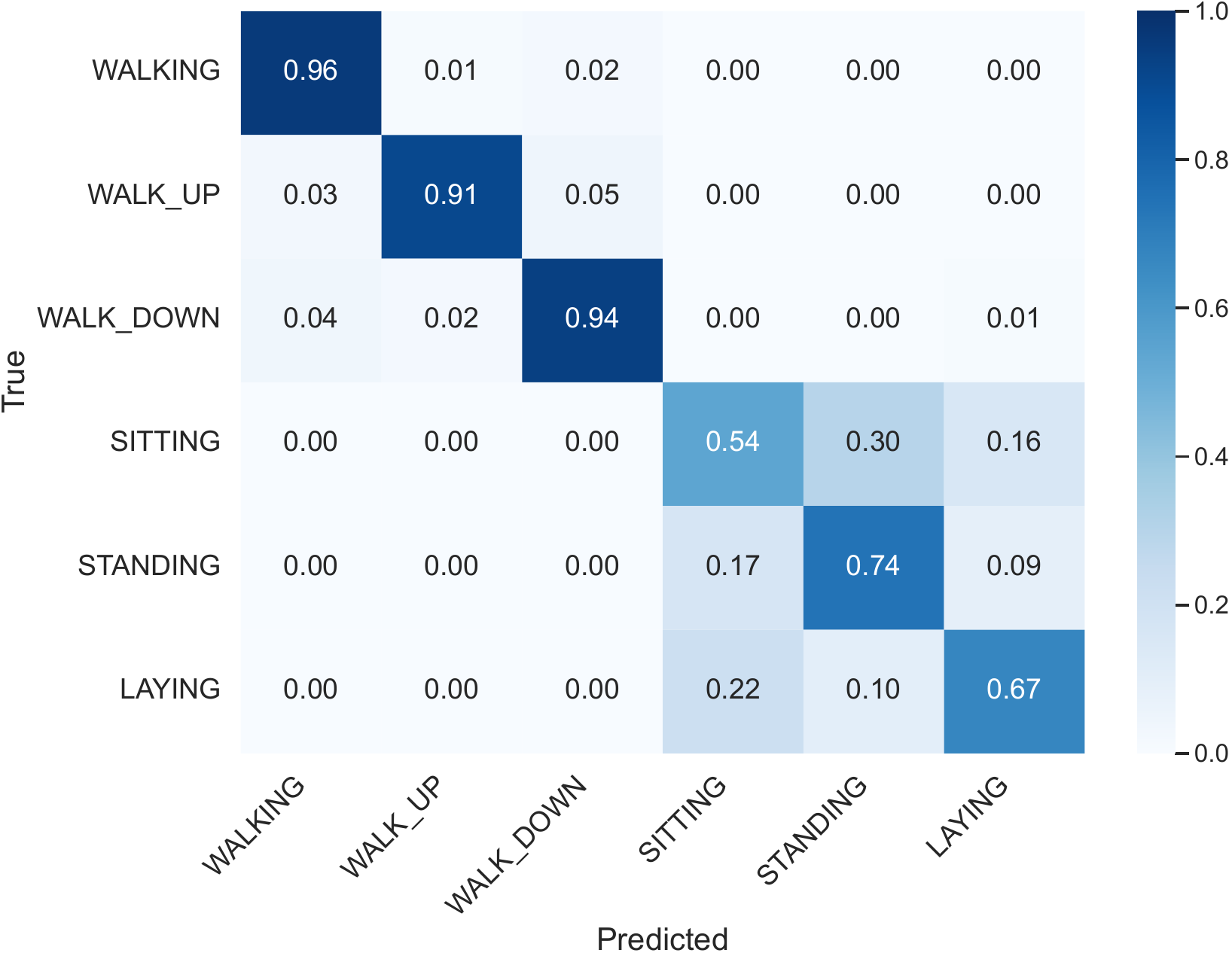}\hfill
  \includegraphics[width=0.49\linewidth]{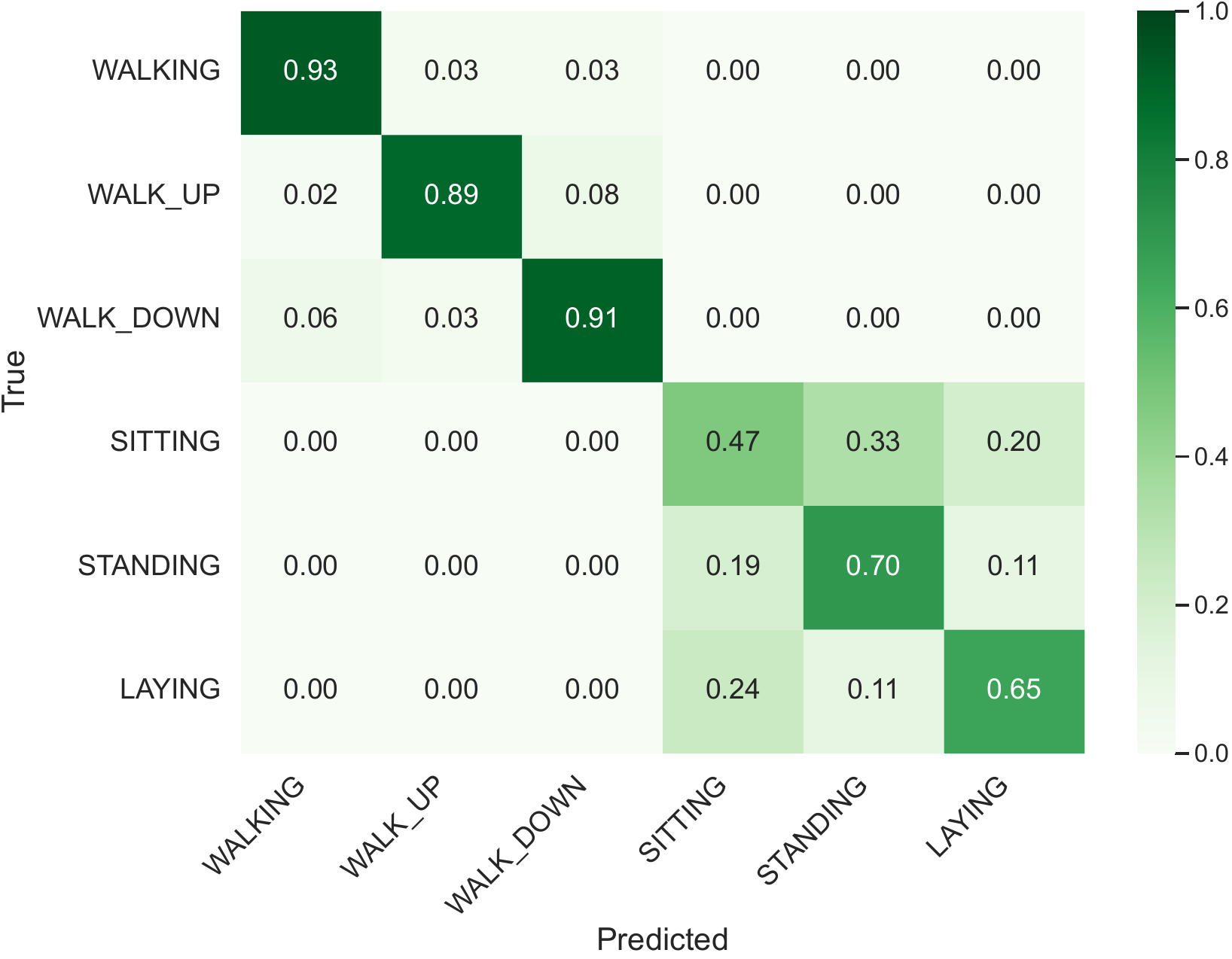}        
  \caption{UCI-HAR analysis. Top: linear-probe performance before and after LoRA. Bottom: confusion matrices (left: NormWear-Base, right: NormWear-LoRA). Degradation concentrates in static
  postures.}   
  \label{app:ucihar_linear_probe}      
\end{figure}

\begin{figure}[!htbp]
  \centering
  \includegraphics[width=0.49\textwidth]{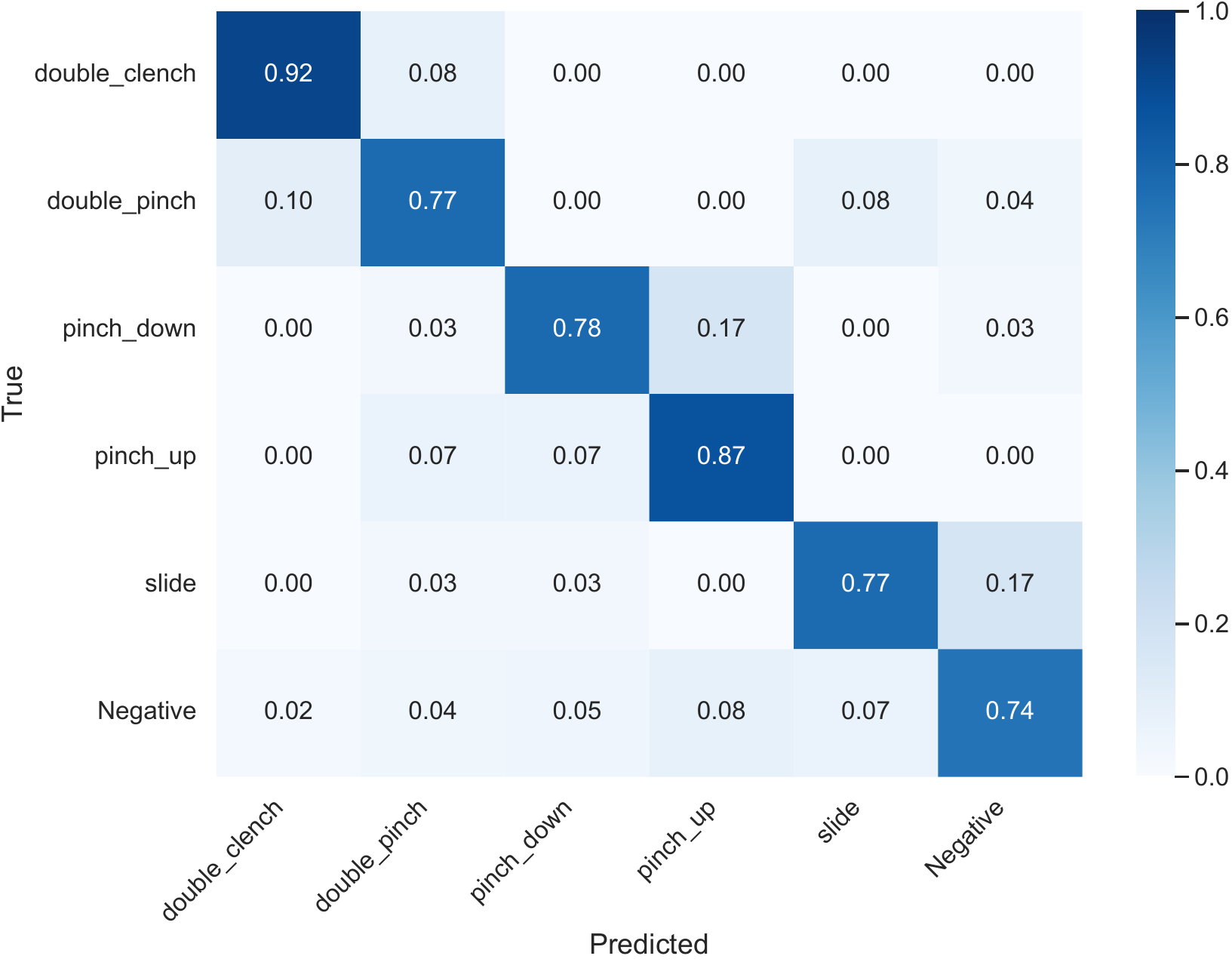}\hfill\includegraphics[width=0.49\textwidth]{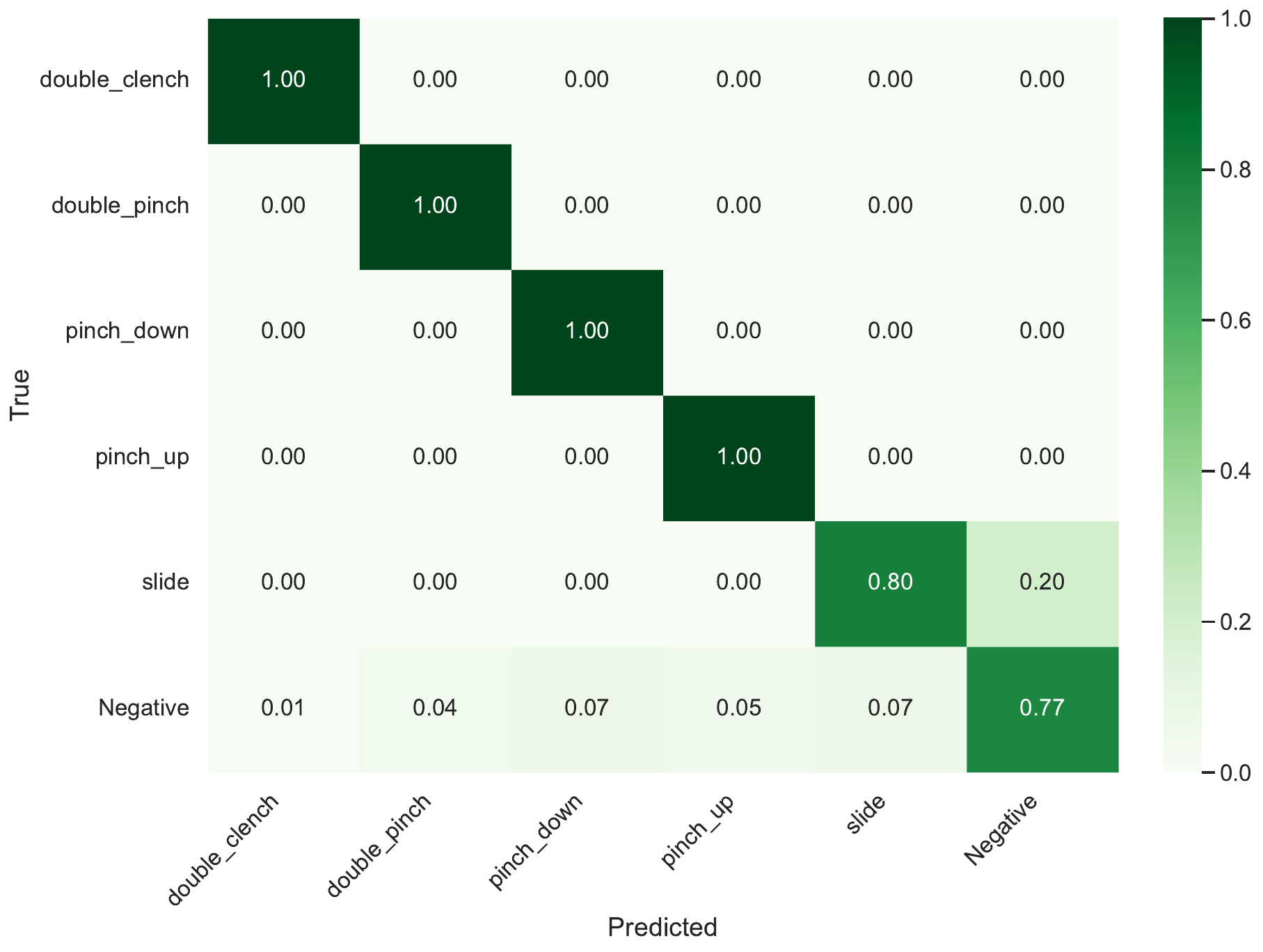}
  \caption{NormWear-LoRA confusion matrices (test set, with augmentation): window-level row-normalized (left) and clip-level row-normalized after aggregation with $k=3$ (right).}
  \label{app:cm_normwear_lora}
\end{figure}

Table~\ref{app:ucihar_linear_probe} reports linear-probe performance on the UCI-HAR dataset before and after LoRA fine-tuning. A performance drop is observed after adaptation, indicating some loss of general-purpose representations. The accompanying confusion matrices reveal that this degradation is primarily concentrated in static activity classes. The row-normalized confusion matrices in Figure~\ref{app:cm_normwear_lora}  at both window and clip levels show that aggregation at the clip level improves classification consistency and reduces misclassification noise.

% ● \begin{wrapfigure}[18]{r}{0.5\textwidth}       
%   \centering
%   \renewcommand{\arraystretch}{1.2}           
%   \small                                  
%   \begin{tabular}{@{} l cccc @{}}
%   \toprule                                                                                                                                                                           
%   \textbf{Model} & \textbf{Acc} & \textbf{F1 (Ma.)} & \textbf{F1 (Wt.)} & \textbf{AUC} \\
%   \midrule NormWear-Base & 0.787 & 0.793 & 0.786 & 0.957 \\                                                                           NormWear-LoRA & 0.750 & 0.755 & 0.749 & 0.948 \\ 
%   \midrule   $\Delta$ & $-0.037$ & $-0.038$ & $-0.038$ & $-0.010$ \\   
%   \bottomrule                                                                                                                                                                        
%   \end{tabular}                                                                                                                                                                                       
%   \end{wrapfigure}   

\end{document}